# CLOUD DETECTION THROUGH WAVELET TRANSFORMS IN MACHINE LEARNING AND DEEP LEARNING

*Philippe Reiter*

Department of Electronic and Electrical Engineering, University of Strathclyde
University of Strathclyde, 16 Richmond St, G1 1XQ, Glasgow, United Kingdom

## ABSTRACT

*Cloud detection is a specialised application of image recognition and object detection using remotely sensed data. The task presents a number of challenges, including analysing images obtained in visible, infrared and multi-spectral frequencies, usually without ground truth data for comparison. Moreover, machine learning and deep learning (MLDL) algorithms applied to this task are required to be computationally efficient, as they are typically deployed in low-power devices and called to operate in real-time.*

*The following paper explains Wavelet Transform (WT) theory, comparing it to more widely used image and signal processing transforms, and explores the use of WT as a powerful signal compressor and feature extractor for MLDL classifiers.*

## 1. INTRODUCTION

Cloud detection is an important component in remote sensing applications. The presence of clouds in remotely sensed images can obscure ground details of interest and invalidate results. Detecting clouds can be a time- and resource-consuming procedure, however, which usually does not have properly annotated datasets for the large variety of cloud types, shapes and altitudes of formation.

The current common practice is to compare a remotely sensed area of interest across a time series of samples and remove cloud cover through an object subtraction process [1]. This is a computationally and time expensive task, though, which can be improved by only capturing images with minimal to no cloud cover. To capture and process images for the presence of clouds in real-time requires efficient cloud detection methods.

Machine learning and neural network algorithms are used to analyse images across a number of spectra, including visible range, infrared and multi-spectral. Given the large variety of cloud types, shapes and altitudes of formation, ground-truth comparison data is often unavailable for supervised classification techniques. Instead, the algorithms rely on precise, low-overhead, automatic feature extraction to feed into unsupervised learning networks [2].

Feature extraction and compression through the application of Wavelet Transforms (WTs) has been researched. The light-weight feature extraction and lossless compression afforded by WTs render these algorithms suitable for implementations in both unsupervised and supervised, where applicable, learning models.

The following sections begin by explaining the theory underlying WTs, and contrast them to more widely deployed image and signal processing transforms. The application of WTs as feature extractors and compressors in both machine learning and deep learning (MLDL) models is then described for Support Vector Machines, Probability Neural Networks, Kohonen Self-Organising Maps, and Convolutional Neural Networks.

It is hypothesised through the research examined that the use of WTs is a powerful feature extraction and compression technique for cloud detection and further MLDL applications.

## 2. IMAGE AND SIGNAL TRANSFORMATIONS

A key concept for the processing of signals and images is the use of transform domains. These mathematical constructs consist of fundamental or basis functions that map complex signals and spaces into atomic building blocks, which can then be more readily processed by digital systems and stored [3].

### 2.1 Discrete Fourier and Cosine Transforms

The Fourier Transform (FT) is a widely used transform domain in numerous digital signal processing (DSP) applications. The FT, or Discrete FT (DFT), when operating on discretely sampled signals, performs a time-to-frequency domain transformation by converting a signal into a sum of sinusoids at different frequencies. It is a powerful transform that is central to DSP, but it presents a drawback for signals whose frequencies change over time – known as non-stationary signals. Whereas the set of frequencies in a stationary, unchanging signal can be computed using the DFT (and even more quickly computed using the Fast Fourier Transform, FFT, technique), for signals whose frequency characteristics change over time, it is not possible to directly use this transform to isolate the precise moments when frequency events occur. One method to capture both time and frequency is discussed in the next subsection, Short-Time Fourier Transform.

Another often used transform in DSP, especially as it regards audio data compression, is the Discrete Cosine Trans-form (DCT). Similar to the DFT, the DCT distils a signal to the sum of correlations with cosine basis functions at different frequencies. Unlike the DFT, which outputs complex number results, or coefficients, DCT coefficients are values in the real numbers. It is an effective transform for so-called "smooth", continuous signals, but is not suitable for signals that contain discontinuities or abrupt changes [3]. As will be explained, an advantage of the Wavelet Transform is its ability to process signals with discontinuities and singularities.

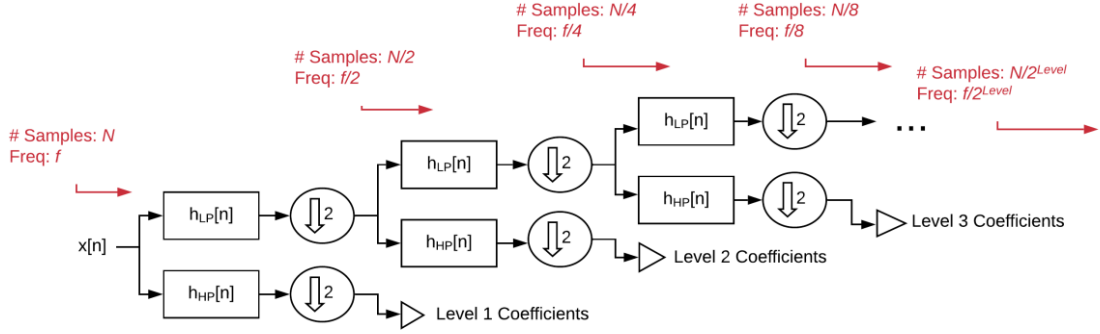

Figure 1 – DWT binary cascade filter bank. Composite diagram from [3][4].

## 2.2 Short-Time Fourier Transform

The Short-Time Fourier Transform (STFT) is intended to solve the inability of the DFT to capture both the time and frequency dimensions of a non-stationary signal. It performs this by dividing a signal into a sequence of windows and individually taking the DFT of each of these windows. Typically, there will be an overlap of windows to ensure minimal information loss [3].

The result of the STFT operation is a representation of a signal in both the time and frequency domains, which can be depicted as a two-dimensional (2D) representation called a spectrogram. Depending on the window size used to compute the transforms, either greater precision in the time domain or in the frequency resolution can be achieved, but not both simultaneously. This trade-off between time and frequency precision is addressed by the Wavelet Transform, which is the subject of the next section.

## 3. WAVELET TRANSFORM

The Wavelet Transform (WT) provides a solution for both the time versus frequency resolution trade-off of the STFT, as well as effectively operating with signals featuring discontinuities and singularities. Both time and frequency precision are attained through the use of windows with varying sizes [3]. Edges and discontinuities appear as high-valued coefficients through the WT [5], which is useful for feature extraction in machine learning and deep learning (MLDL) contexts, as will be explored later in this paper.

Basis functions for the WT are frequency-scaled and time-shifted versions of the "mother wavelet" [3], a finite-energy wave in both the time and frequency domains [1]. The mother wavelet and WT equations follow from [3]:

$$\textit{Mother Wavelet}: \varphi_{s,\tau}(t) = \frac{1}{\sqrt{s}} \varphi\left(\frac{t-\tau}{s}\right) \quad (1)$$

where $s$ is the frequency scaling factor and $\tau$ is the time dimension

$$\textit{WT}: \gamma(s,\tau) = \int x(t)\varphi^*_{s,\tau}(t)dt \quad (2)$$

where $x(t)$ is the signal on which to perform the WT, and $(s, \tau)$ are the WT coefficients

As can be deduced from the mother wavelet equation (1), there is an inverse relationship between the scaling factor, $s$, and the waveform shape. Larger values of $s$ compress the waveform and result in narrowband windows for frequency precision, while smaller values stretch the waveform to have a wideband spectrum for time sensitivity [3]. This frequency scaling enables the WT to capture both time and frequency information with precision.

In addition, the WT equation (2) can be shown to be linearly independent and invertible [3]. The sparsity of the resulting transform matrix lends to the use of the WT for dimensionality reduction [6]. The benefit of these properties will be further expounded in the Machine Learning and Deep Learning Applications section.

## 3.1 Continuous and Discrete Wavelet Transforms

The previous section introduced the WT as well-capturing both time and frequency components of a continuous, possibly non-stationary signal. Although not thoroughly explored in this paper, the WT depends on the family of wavelets used, such as Haar, Daubechies, and Discrete Meyer [1]. The Haar wavelet is regarded as the simplest wavelet and least computationally demanding [7]. Daubechies (db) wavelets with varying threshold, or vanishing point, levels are orthogonal, compact wavelets well-matched for image analysis and compression [3]. Figure 2 depicts a Daubechies wavelet with threshold level of 4, db4, as plotted with MATLAB.

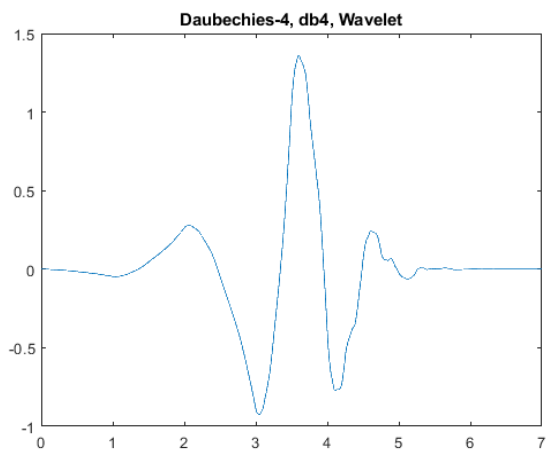

Figure 2 – Daubechies wavelet with threshold of 4, db4 [8].

The Discrete Wavelet Transform (DWT) represents the continuous WT with discrete scaling and time. It is the DWT, specifically, that can be substituted for the STFT, and supports both time and frequency transformation without the resolution trade-off that affects the STFT.

The DWT can be conceptualised as a cascaded filter bank [3], with applications in signal processing, feature extraction and, as previously mentioned, data compression [9]. Figure 1 diagrams three levels of a DWT filter bank, with the input signal, x[n], initially being divided between low-pass (LP) and high-pass (HP) filters. To account for scaling, the filtered signals are downsampled by a factor of two at the end of each processing level [3]. The DWT coefficients are extracted from the HP filter stage post decimation. Increasingly finer details are extracted following each sequential filtering level [9].

For images, a 2D DWT is performed by computing 1D DWT operations first across all columns, then all rows [3].

### 3.2 Wavelet Scattering Transform

The Wavelet Scattering Transform (WST) is a variant of the DWT intended for signals with slight changes and variations through time. This transform showcases properties that are well-aligned with the requirements for MLDL; namely, it is invariant to signal or image rotation and translation, it supports signal discontinuities and singularities, as previously mentioned, and its data representation is compact [6], making it valuable for complex MLDL networks.

Each scale, or scattering path [9], represents a different form of the image or signal. This is similar to the process of image augmentation through scaling and rotation applied for conventional artificial neural networks (ANNs). Just as image augmentation is typically leveraged to boost the number of distinct samples in a limited training set, so too can WST scattering be used to augment a dataset for improved feature extraction [7].

As detailed, the DWT and WST can be used for feature extraction, dimensionality reduction [2], as well as image compression [1] and augmentation, which form the core of most MLDL implementations. The features extracted can then be passed to either a shallow or deep regression or classification flow using a variety of machine learning algorithms, ANNs and deep neural networks (DNNs) [6][9].

### 4. MACHINE LEARNING AND DEEP LEARNING APPLICATIONS

A number of independent research papers were studied for uses of WT as a feature extractor for MLDL classification and regression. Common networks, as they pertain to cloud detection, are described in the following subsections.

### 4.1 Shallow Learners and Artificial Neural Networks

A Kohonen Neural Network (KNN) [2][10] and a Probability Neural Network (PNN) [7] have been shown to be effective as classifiers post WT feature extraction. A PNN is similar to a Radial Basis Function (RBF) neural network, which requires only a single pass-through of all the data, making it a rapid learning implementation [7].

A KNN, or Kohonen Self-Organising Map (SOM), is particularly interesting for the purposes of cloud detection, as it is an unsupervised algorithm [10]. Since the volume of cloud properties and characteristics is vast, the ability of a KNN to automatically learn patterns and similarities in cloud image-sets is advantageous. This is achieved by the Kohonen algorithm, which clusters data in such a way as to minimise the differences between samples within a cluster and maximise the differences between clusters.

Neural networks (NNs) are complex and computationally expensive, however. In cases where a light-weight algorithm is required, one research paper [11] has shown a Support Vector Machine (SVM) classifier to work well with the outputted features from WTs. Instead of the NN method of detecting patterns through one or more neuron layers, an SVM classifier compartmentalises data through a set of tuned hyperplanes. This segmentation through hyperplanes is significantly quicker and less computationally taxing than NNs, so coupling a WT with an SVM is a desirable implementation for appropriate datasets that require minimal resources.

### 4.2 Multi-Level Wavelet CNN

Convolutional Network Networks (CNNs) have become increasingly popular in MLDL applications, especially in the areas of computer vision and object detection. A Pooling layer in a CNN performs downsampling of feature maps to maintain sufficiently sizeable representations, or receptive fields, further into the network while reducing the overall computational requirements. Downsampling is performed by condensing each region of a feature map to a single value, corresponding to either the average or maximum value found in the associated region. Pooling is a destructive operation, and results in the loss of information, with the trade-off being computational efficiency.

In [12], a Multi-Level Wavelet CNN (MWCNN) is proposed, which substitutes the conventional Pooling layer in a

CNN with a DWT layer. The DWT performs a similar operation to Average Pooling, but does not result in information loss. This powerful advantage comes about from the invertibility of the DWT, as explained in the Continuous and Discrete Wavelet Transforms subsection, above.

Further research on this topic needs to be conducted to confirm that the performance and information integrity of a DWT layer in an MWCNN is superior to that of Pooling in CNNs across numerous applications.

## 5. CONCLUSION

The capability to capture both time and frequency precision, using variable window sizing, advantages WT functions over the more widely used DCT and STFT transforms for image and signal analysis and processing. Furthermore, lossless image reconstruction can be achieved due to the invertibility of the DWT.

WTs have been explored as functions well-suited for MLDL applications. The feature augmentation, dimensionality reduction, and lossless compression of the WST makes this function particularly applicable to deep neural networks as a replacement for and enhancement over an Average Pooling layer. Software libraries, such as *ScatNet* for MATLAB and *Kymatio* for Python [13], have already been developed to ease the effort by programmers to leverage WST in their MLDL networks. By performing feature extraction, compression and dimensionality reduction through WTs, greater efficiency and potentially improved results can be achieved in cloud detection and further MLDL applications.